\documentstyle[pre,aps,multicol,epsfig]{revtex}
\epsfclipon

\begin{document}   

\tightenlines

\title{Comment on ``Deterministic equations of motion and phase ordering
dynamics''}
\author{Julien Kockelkoren and Hugues Chat\'e}
\address{CEA --- Service de Physique de l'Etat Condens\'e, 
Centre d'Etudes de Saclay, 91191 Gif-sur-Yvette, France}
\maketitle
\begin{abstract}
Zheng [Phys. Rev. E {\bf 61}, 153 (2000)] claims that phase ordering dynamics in
the microcanonical $\phi^4$ model displays unusual scaling laws. We show here,
performing more careful numerical investigations,
that Zheng only observed transient dynamics mostly due to the corrections
to scaling introduced by lattice effects, and that Ising-like (model~A)
phase ordering actually takes place at late times. Moreover, we argue that
energy conservation manifests itself in different corrections to scaling. 
\end{abstract}

\begin{multicols}{2} 

\narrowtext

The problem of the dynamical foundations of statistical mechanics 
has received renewed attention recently, when, in the spirit of the 
famous work of Fermi, Pasta, and Ulam \cite{FPU}, researchers  directly
studied the evolution
of isolated, many-degrees-of-freedom Hamiltonian systems with the
aim of relating their microscopic, deterministic, chaotic motion 
to their macroscopic statistical properties \cite{CAIANI2}. 
In this context, the two-dimensional lattice $\phi^4$ model is of special
interest  because it is known to exhibit, within the canonical
ensemble, a second-order phase transition in the Ising universality class.
Recent work \cite{CAIANI1,BZ-PRL} aimed, 
in particular, at studying the corresponding behavior 
in the isolated, microcanonical case, whose equations of motion can
be written:
$$
\ddot{\phi_i} = \sum_j (\phi_j-\phi_i) + m^2 \phi_i -\frac{g}{6} \phi_i^3
$$
where the sum is over the four nearest neighbors of site $i$ on a square lattice.

In the same spirit, Zheng \cite{BZ} has considered the a priori 
simpler problem of the phase-ordering process which takes place when 
the microcanonical $\phi^4$ lattice is suddenly ``quenched'' 
below the critical point. Universality of domain growth laws is
nowadays a fairly well-established topic \cite{BRAY}. It is well documented,
even for less traditional systems such as deterministic, possibly
chaotic, spatially-extended dynamical systems \cite{CMLPER,DDAP,APS,STRA-COM}. 
For a scalar order
parameter, two main universality classes
can be distinguished depending on whether it
is locally conserved or not. The Ising model is prototypical of the 
non-conserved case, and so should be the lattice $\phi^4$ model,
at least in the usual canonical ensemble point of view. However,
and this was the interesting point raised by Zheng, the presence of 
the energy conservation in the microcanonical case might have an influence
on the dynamical scaling laws associated with phase ordering. 
In this sense, the question is whether the phase ordering of model C
\cite{HOHA} is in the same universality class as model~A. 

Using numerical simulations in the so-called ``early-time'' regime,
Zheng confirmed that the usual dynamical scaling laws seem to hold,
but with exponents at odds with those both the non-conserved order parameter
(NCOP) and the conserved order parameter (COP) class \cite{BZ}.  
In particular, he found  $z=2.6(1)$ ( $1/z$ is the exponent governing the
algebraic growth of $L$, the typical size of domains), between its
NCOP and COP values ($z_{\rm NCOP}=2$ and $z_{\rm COP}=3$).
Zheng ``explains'' this surprising result by the influence on scaling
of ``the fixed point corresponding to the minimum energy of random initial
states'' (present when considering domain growth).

Here we show that a more careful numerical 
investigation actually leads to conclude that the microcanonical
lattice $\phi^4$ model shows normal, NCOP phase-ordering. We argue that
energy-conservation, but also lattice effects, have sub-leading influences
on scaling. We discuss in particular the effect of the increase of the
``bulk temperature'' due to the progressive disappearance of interfaces
during the growth transient. We attribute the erroneous results of Zheng 
to the danger of using ``early-time'' methods and naive logarithmic plots 
in problems with large microscopic times and/or corrections to scaling.

\section{Normal scaling at late times}

Zheng conducted different numerical experiments which led to
estimates of exponents $z$, $\lambda/z$ (where $\lambda$ is 
the so-called Fisher-Huse exponent),
and $\theta=(2-\lambda)/z$. From all direct measurements of $z$ he concluded
that $z=2.6(1)$. Using this value, he found an estimate of $\lambda$
in agreement with the NCOP value ($\lambda=1.22(5)$ whereas 
$\lambda_{\rm NCOP}=5/4$). Thus, the only strong departure from the NCOP
values is for exponent $z$. Therefore, in the following, we focus on
growth law of $L$ in large systems at late times (as opposed to
the early-time approach favored by Zheng). 

In order to reach late times satisfactorily, 
we need a better control of the conservation of energy
than with the simple second-order scheme used by Zheng. 
The following results were obtained with a third-order bilateral
symplectic algorithm \cite{CASETTI} with a timestep $dt=0.025$.
The conservation of initial energy is better than $10^{-5}$ in relative value
in all runs presented. 
To investigate phase-ordering, we use the same initial conditions as Zheng 
($\phi=\pm q_0$ where the sign is random and $q_0$ calculated to
yield the desired energy density $\epsilon$). 
We present results for two sets of parameter values, $(m^2,g)=(6,1.8)$
(set A, used by Zheng), and  $(m^2,g)=(2,0.6)$ (set B, used in 
\cite{CAIANI1,BZ-PRL}). The initial energy density
($\epsilon_0 = 27$ for set A, and 
$\epsilon_0 = 10.0001$ for set B),
was chosen very close
to its minimum value allowed by the random-sign initial conditions
($\epsilon_{\rm min} = 80/3$ for set A, and 
$\epsilon_{\rm min} = 10$ for set B). This ensures that 
``thermal'' fluctuations are minimized, since the energy density is then
as far as possible from
the critical energy density $\epsilon_{\rm c}$
($\epsilon_{\rm c}\approx 35$ for set A, and $\epsilon_{\rm c}\approx 21$ 
for set B).


\begin{figure}
\centerline{
\epsfxsize=86mm
\epsffile{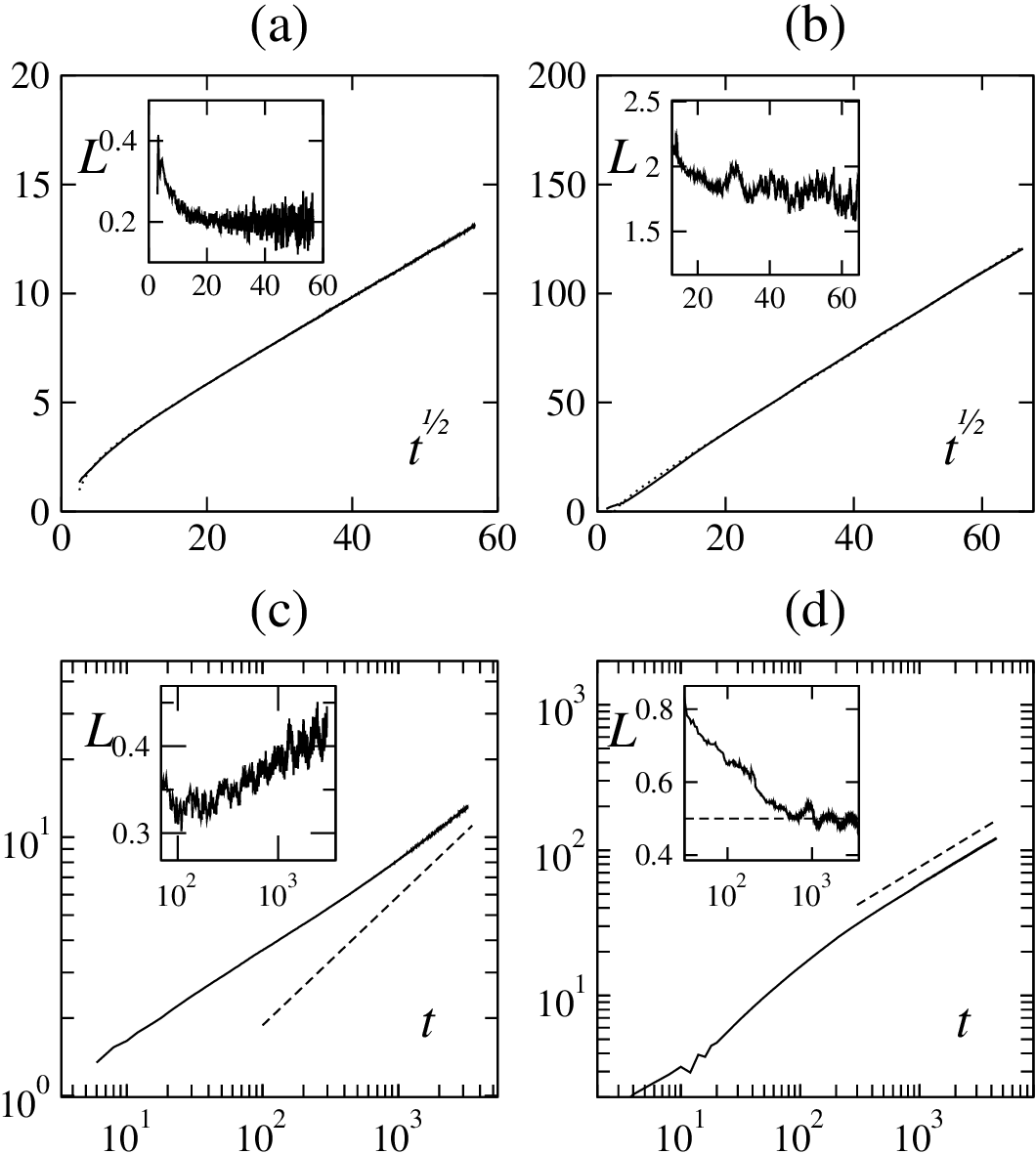}
}
\caption{Growth of typical scale $L(t)$.
All results presented here were obtained from single runs of lattices
of $8192\times 8192$ sites with periodic boundary conditions.
(a,b): $L$ vs $\sqrt{t}$ for parameter set A and B, the dotted lines
are the fits discussed in section \ref{fits};
(c,d): same but in logarithmic scales (dashed lines for ``expected'' 
behavior $z=2$).
Insets: local slopes calculated by a running average of the derivated 
signal over a window of fixed size in the corresponding variable 
(i.e. $\sqrt{t}$ or $\log_{10}t$ here). 
For (a,b) window size $\Delta(\sqrt{t})=0.5$; 
For (c,d) window size $\Delta(\log_{10}{t})=0.05$.
}
\label{f1}
\end{figure}



\subsection{Growth of typical domain size}

The typical domain size $L$ was determined by the mid-height value of 
$C(r,t)$, the normalized two-point correlation function calculated
for simplicity along the principal axes of the lattice using the continuous
field $\phi$ or the reduced ``spin'' variable $\sigma\equiv {\rm sign}(\phi)$:
$C(k,t)=\frac{1}{2N} \sum_{ij} \sigma_{i j} (\sigma_{i+k\ j} +
\sigma_{i\ j+k})$. 
(No significant difference was found between the two cases, and only results
using $\sigma$ are shown below.)
We first checked that dynamical scaling holds by observing the
collapse of $C(r,t)=f(r/L(t))$ curves at different times 
after some transient (not shown) \cite{note1}.

For both sets of parameters, the expected NCOP law ($L \sim \sqrt{t}$,
i.e. $z=2$)
is reached at late times, but a rather long transient is observed,
especially for set~A (Fig.~\ref{f1}ab). The same data plotted in 
logarithmic scales is thus misleading. If the data for set B reach
the ``normal'' scaling (see Fig.~\ref{f1}d and its inset) at late times,
the corresponding plot for parameter set A (Fig.~\ref{f1}c) seems to indicate
a value of $1/z$ between $\frac{1}{3}$ and $\frac{1}{2}$ (typical of the 
value estimated by Zheng)
if one ignores the systematic trend upward of the local exponent 
(see inset of Fig.~\ref{f1}c) \cite{note2}.

\begin{figure}
\centerline{
\epsfxsize=86mm
\epsffile{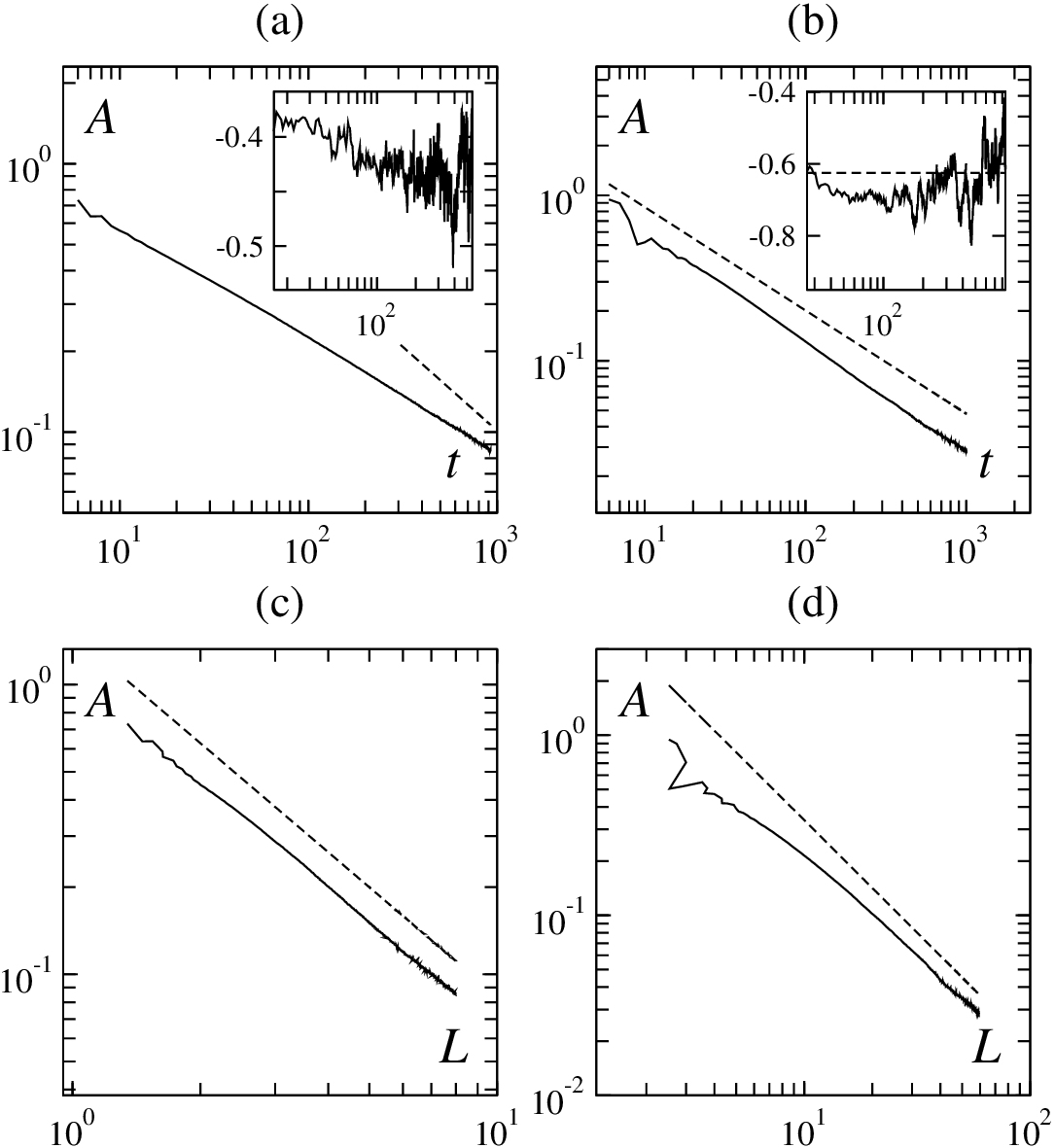}
}
\caption{Decay of autocorrelation function $A(t,t_0)$ from single runs
of lattices of $4096 \times 4096$ as in
Fig.~\protect\ref{f1} ($t_0=5$). 
(a,b): $A$ vs $t$ for parameter sets A and B  
(logarithmic scales, dashed lines: slope $\lambda/z=5/8$, insets: local slopes);
(c,d):  $A$ vs $L$ (logarithmic scales) for parameter sets A and B 
(logarithmic scales, dashed lines: slope $\lambda=5/4$, 
the quality of the data is not high enough for local slopes).
}
\label{f2}
\end{figure}

\subsection{Decay of autocorrelation function}

Following Zheng, we also measured the decay of the autocorrelation function
$A(t,t_0)=\langle \sigma(t) \sigma(t_0) \rangle$, but,
again, for rather large systems and late times,
choosing, in particular, initial reference times $t_0$ larger than the
``microscopic'' transient time. In the usual NCOP phase ordering framework,
we expect
$$
A(t,t_0) \sim \left[\frac{L(t_0)}{L(t)}\right]^{\lambda} \sim 
\left(\frac{t_0}{t}\right)^{\lambda/z} 
$$
where the Fisher-Huse exponent $\lambda=5/4$. Again, plotting $\log A$
vs $\log t$ for parameter set A may yields an ``effective'' exponent
$\lambda/z$ smaller than its NCOP value (and close to the value given 
by Zheng), but a closer look reveals 
a systematic increase of the instantaneous exponent 
(Fig.~\ref{f2}a and inset). However,
plotting $A$ vs $L$, the expected scaling is observed
(Fig.~\ref{f2}c). Considering now parameter set B, the expected scaling laws
are observed rather easily (Fig.~\ref{f2}b,d).
This confirms further  that for
the parameter values chosen by Zheng, the onset of the asymptotic 
scaling regime is delayed.

\begin{figure}
\centerline{
\epsfxsize=86mm
\epsffile{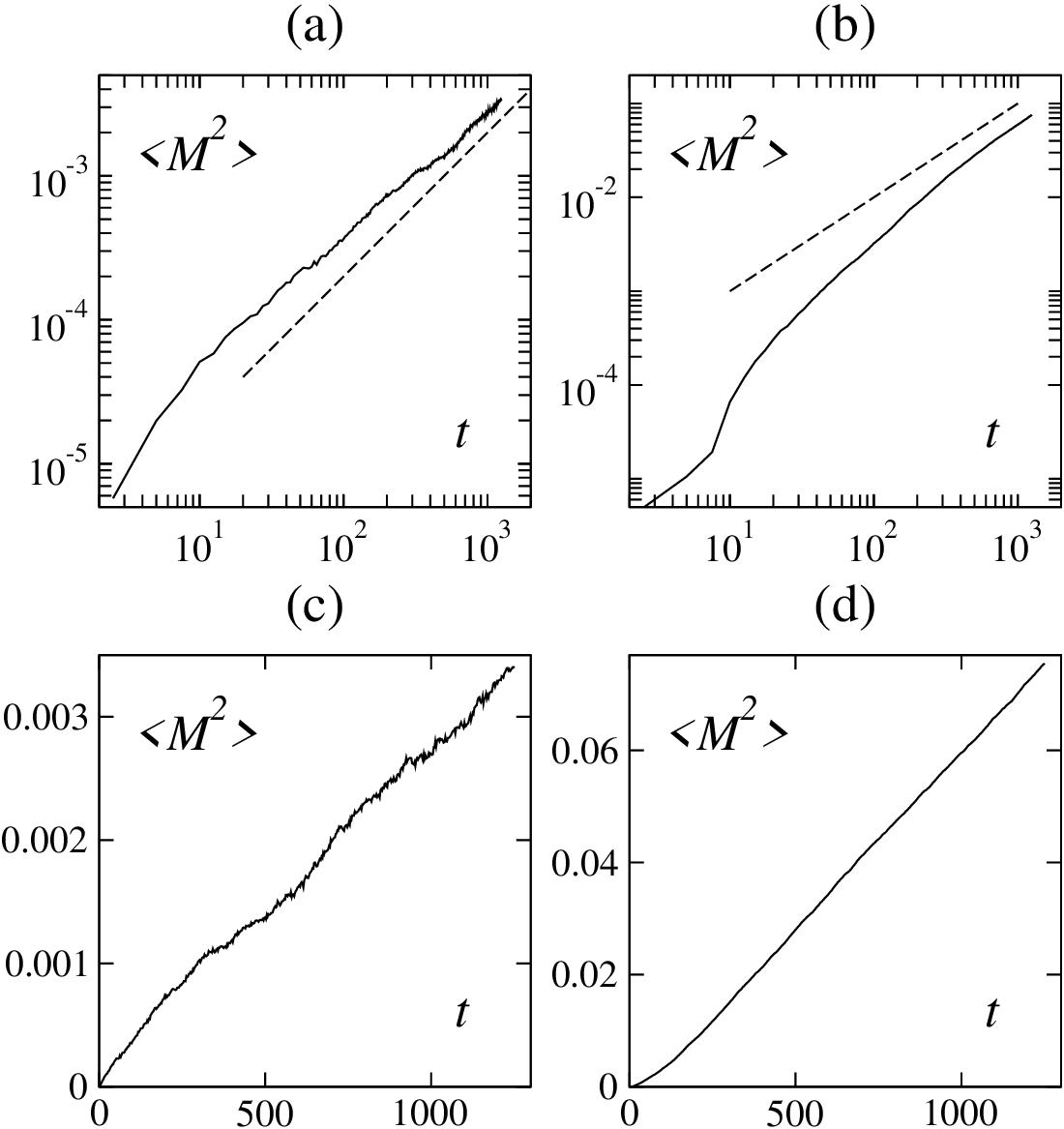}
}
\caption{Early-time growth of the squared magnetization for an ensemble of
500 symmetric ($M=0$) samples in a lattice of $512\times 512$ sites.
(a,b): $\langle M^2 \rangle$ vs $t$ (logarithmic scales) 
for parameter sets A and B;
dashed lines: slope 1.
(c,d):  $\langle M^2 \rangle$ vs $t$ (linear scales) 
for parameter sets A and B.}
\label{f3}
\end{figure}

\subsection{Early-time growth of squared magnetization}

For the sake of completeness, and in order to probe the validity of 
the early-time scaling approach taken by Zheng, we also performed 
numerical simulations to measure the short-time growth of the squared 
magnetization $M^2$
($M$ being defined as the spatial average of either $\phi_i$ or 
$\sigma_i = {\rm sign} (\phi_i)$)
for samples with zero initial magnetization. In this case, for NCOP
scaling, we expect $\langle M^2\rangle \sim t^{d/z} = t$.
Our data on logarithmic scales barely reaches the expected behavior
at $t\simeq 10^3$ (Fig.~\ref{f3}a,b). Note that the corrections 
have different sign for the two parameter sets. On linear scales,
however, our data reveals the expected proportionality of $\langle M^2\rangle$
and $t$ (Fig.~\ref{f3}c,d).

\section{Corrections to scaling: energy conservation and lattice effects}

The above results show that domain growth in the microcanonical $\phi^4$
model eventually falls into the NCOP universality class ($z=2$, $\lambda=5/4$)
after some possibly long transient behavior. In this section, we suggest that
there are two main factors at the origin of these corrections to scaling:
space discretization and energy conservation.

\subsection{Lattice effects}

For the sets of parameters studied by Zheng (notably parameter set A),
domain growth initially appears to be slower than the expected 
NCOP law
(the effective value of $z$ measured at short times is larger than two).

This is similar to earlier observations on coupled map lattices,
both for the NCOP and COP cases. First measurements of domain
growth seemed to indicate slower growth in those discrete-space,
discrete-time chaotic models \cite{CMLPER,APS},
but it was shown later that this non-trivial
scaling is only transient and that, for late-enough times, normal
scaling is recovered \cite{DDAP,STRA-COM}. 
It was also argued that these long transients
gradually disappear in the continuous-space limit which is well-defined
in these systems \cite{CMLRG}.

Here, it is easy to observe that, increasing $g$ (and $m^2$, since there
is only one free parameter), the slowing-down
of domain growth due to lattice effects diminishes, as suggested by the
difference between parameter sets A and B. For parameter set B, the
transient actually presents faster growth than asymptotically: corrections 
to scaling are dominated by another phenomenon, stronger than lattice effects.

\begin{figure}
\centerline{
\epsfxsize=86mm
\epsffile{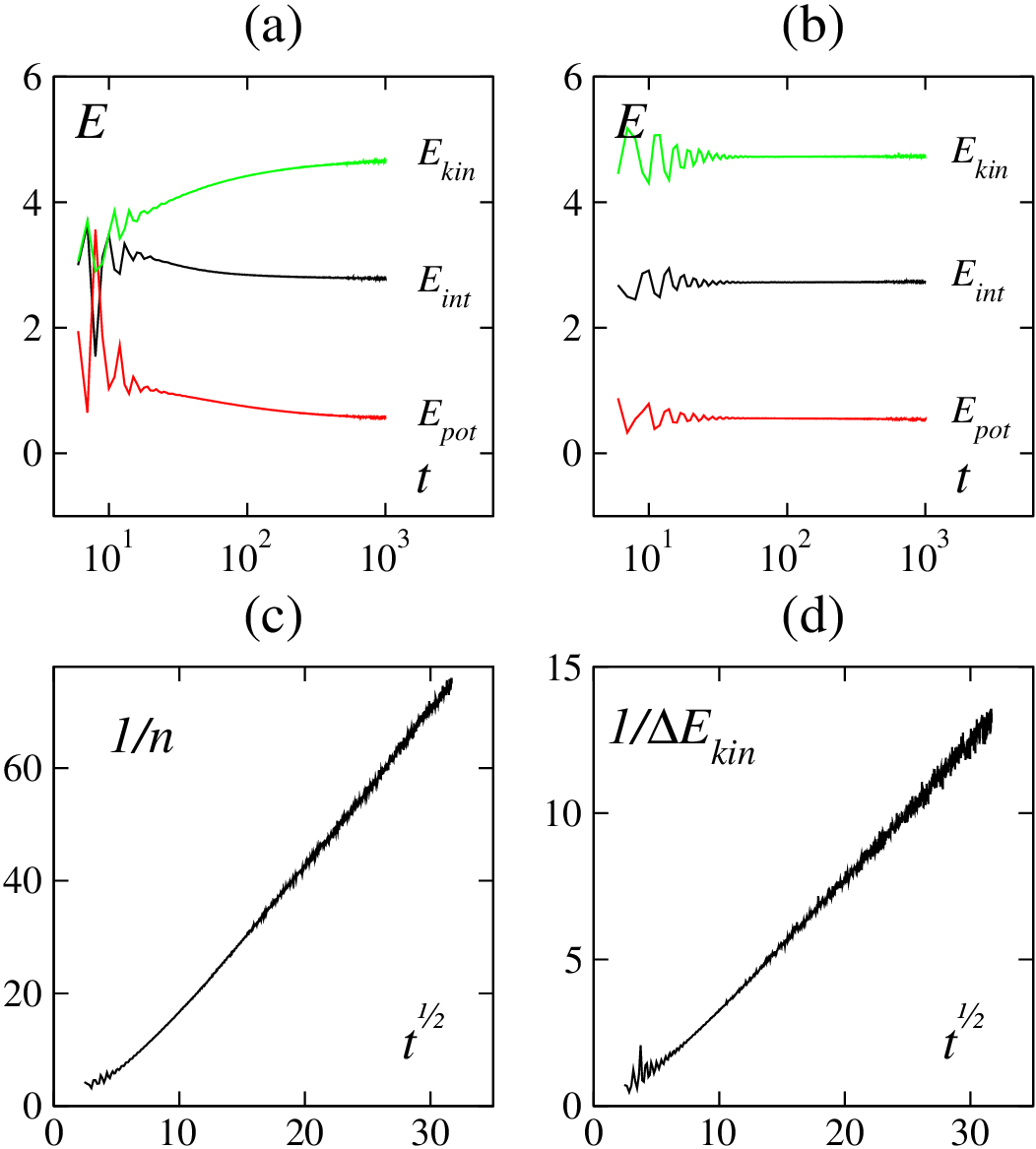}
}
\caption{Corrections to scaling due to energy conservation. 
Results for parameter set B, for which these corrections are dominant 
(see text).
(a): the three components of the total energy during phase ordering.
(b): same but from ordered initial conditions.
(c): decay on the number of interfacial sites (defined as sites with a least
one nearest-neighbor of opposite sign); the number of interfacial sites
measured during the evolution from an initially-homogeneous configuration
has been substracted.
(d): Convergence of the total kinetic energy (temperature) towards its
asymptotic value: $1/\Delta E_{\rm kin}$ vs $\sqrt{t}$ during phase ordering.
}
\label{f4}
\end{figure}

\subsection{Energy fluxes}

The (conserved) total energy in the system can be partitioned into
three components, kinetic, potential, and interaction energy: 
\begin{eqnarray*}
E_{\rm kin} & = & \sum_i \frac12 \dot{\phi_i}^2  \\
E_{\rm pot} & = & \sum_i - \frac12 m^2 \phi_i^2 + \frac{g}{4!}
\phi_i^4 \\ 
E_{\rm int} & = & \sum_i \frac{1}{2} \sum_{j=1}^d (\phi_{i+j}
- \phi_i)^2 \;.  \nonumber
\end{eqnarray*}
As domain growth proceeds following random initial conditions, the interaction
energy decreases with time as a large part of it is contained in the
interfaces separating domains, the density of which decreases like
$1/\sqrt{t}$. Potential decreases and kinetic energy increases (Fig.~\ref{f4}).

Starting from ``ordered'' initial conditions (all sites in the 
same phase, e.g. $\forall i \;\phi_i(0)>0$), the three components of
the total energy are almost constant, except for
a slight (quasi logarithmic in time) increase of interaction energy
and decrease of potential energy 
due to the nucleation of ``thermal'' droplets and to relaxation further into
potential wells. This effect is indeed 
larger for parameter set A for which the minimal possible energy density
$\epsilon_{\rm min}$
is further away from the critical energy density, leading to
stronger thermal fluctuations than for parameter set B.

\subsection{Corrections to effective temperature}
\label{fits}

Kinetic energy can be interpreted as a temperature in
the microcanonical context \cite{CAIANI1}. 
We can thus see the observed increase of 
kinetic energy as an increase of the temperature of the system. 
For the (canonical) Ising model, it is well known \cite{LACASSE} 
that the prefactor
of the growth law of $L(t)$ decreases to zero as the
temperature approaches its critical value. We suggest to see the growth
law observed here for the $\phi^4$ model as including 
a ``temperature-dependent'' prefactor:
\begin{equation}
L(t) \simeq K(E_{\rm kin})\, \sqrt{t}
\label{growth}
\end{equation}
Quantitatively (Fig.~\ref{f4}), 
the kinetic energy seems to reach its asymptotic value like:
\begin{equation}
\Delta E_{\rm kin} \equiv E_{\rm kin}(t)-E_{\rm kin}(\infty) \sim t^{-1/2}
\label{deltaE}
\end{equation}

Assuming its analyticity, we can write the prefactor $K$:
\begin{equation}
K(E_{\rm kin})=K_\infty + K'_1\, \Delta E_{\rm kin} + K'_2\,(\Delta E_{\rm kin})^2 + \ldots
\label{expand}
\end{equation}
where $K_\infty$ is the asymptotic ($t\to\infty$) 
prefactor of the domain growth law.
Injecting (\ref{deltaE}) and (\ref{expand}) into (\ref{growth}),
we finally expect the following Ansatz to hold for the domain growth law:
\begin{equation}
L(t) = K_\infty \sqrt{t} + K_1 + \frac{K_2}{\sqrt{t}} + \ldots
\label{ansatz}
\end{equation}

Equation~(\ref{ansatz}) provides excellent fits to our data 
for the growth of $L(t)$.  We find
$K_\infty \simeq 0.2$, $K_1\simeq2.0$, and $K_2\simeq-3.5$ 
for parameter set A (with $\chi^2=7.8$ and correlation coefficient 0.99996) , 
and $K_\infty \simeq 2.1$, $K_1\simeq-5.6$, and $K_2\simeq5.7$ (with
$\chi^2=10^5$ and correlation coefficient 0.9997) 
for parameter set B. 

The corrective terms have opposite signs in both cases: this indicates that
Eq.~(\ref{ansatz}) is only relevant for parameter set B, because, in analogy
with the Ising model, only negative values of $K_1$ are allowed (the prefactor
of the growth law decreases with increasing temperature). A first 
conclusion is thus that the main corrections to scaling for parameter set A
(typical of those used by Zheng) are due to lattice effects and are {\it not}
consequences of the conservation of global energy. We note that, similarly,
a fit of domain growth in coupled map lattices also yields positive 
values of $K_1$, indicating that this sign is a signature of lattice
effects \cite{DDAP}.

On the other hand, the above framework does provide the relevant explanation 
for the corrections to scaling observed for parameter set B, which can thus be
traced back to the fluxes between the various components of the energy
induced by the decrease of interfaces between domains 
as phase-ordering proceeds.

\section{Conclusion}

In this Comment, we have shown that Zheng reached erroneous conclusions
when studying phase ordering in the microcanonical lattice $\phi^4$ 
model. This system, like other chaotic, deterministic, dynamical
systems presenting phase ordering, 
does display the expected domain growth scaling laws,
i.e. those of the non-conserved order parameter case (model~A, 
$z=2$, $\lambda=5/4$).
We have shown further that the main influence of the conservation of energy
is to introduce corrections to scaling, but that the long transients which
plagued Zheng's approach are due to lattice effects.

Zheng offered, as an explanation for his non-conventional results, that
the system considered falls into the class of model~C,
where a conserved density is coupled to the order parameter \cite{HOHA}.
Studies of phase-ordering in model~C 
\cite{Ohta,Chakrabarti,TBP} show that model~B-like,
but also model~A-like behavior can be observed. In this context, the
microcanonical $\phi^4$ lattice can be considered as a model~C system
quenched into its ``bistable region'' (where $z=2$ is observed
\cite{Elder,TBP}).

\vspace{12pt}

We thank A. Lema\^{\i}tre and A.~S.~Somoza for interesting discussions and
L. Casetti for providing a sample version of his program.

\end{multicols}


\begin{thebibliography}{99}
\bibitem{note1} Since $L$ is precisely defined as the midheight value
of $C(r)$, it is in our opinion not so surprising that we obtain
already a good collapse before the scaling regime sets in. In this
sense our appoach is different from that of Zheng who tried to
collapse $C(r/t^{1/z})$.
\bibitem{note2} Of course, plotting the quantity of
interest against $t^{1/z}$ and looking for a straight line
is not the most accurate way of determining $z$ (here we use this only
because the log-log plots do not exhibit a constant asymptotic slope).
Nevertheless, the value $z=2.6$
measured by Zheng can safely be rejected since the local slope of $L$
against $t^{1/2.6}$ increases with time (not shown here).
\end{thebibliography}
\end{document}